\newcommand{\PR}[3]{Phys.\ Rep.\ {\bf #1},\ #2 (#3)}
\newcommand{\PRL}[3]{Phys.\ Rev.\ Lett.\ {\bf #1},\ #2 (#3)}
\newcommand{\RMP}[3]{Rev.\ Mod.\ Phys.\ {\bf #1},\ #2 (#3)}

\newcommand{\SC}[3]{Science\ {\bf #1},\ #2 (#3)}

\newcommand{\PRA}[3]{Phys.\ Rev.\ A\ {\bf #1},\ #2 (#3)}

\newcommand{\PRE}[3]{Phys.\ Rev.\ E\ {\bf #1},\ #2 (#3)}

\newcommand{\JPBB}[3]{J.\ Phys.\ B:\ At.\ Mol.\ Opt.\ Phys.\ {\bf #1},\ #2 (#3)}
\newcommand{\JPAA}[3]{J.\ Phys.\ A:\ Math.\ Gen.\ {\bf #1},\ #2 (#3)}

\newcommand{\GP}{Gross-Pitaevskii~}

\documentclass[aps,pra,superscriptaddress,twocolumn,secnumarabic,showpacs,amsmath,amsfonts,amssymb,floatfix]{revtex4-1}

\usepackage{amsmath}
\usepackage{verbatim}
\usepackage{graphicx}
\usepackage{subfigure}
\begin{document}

\title{Topology of Quantum Grey Soliton in Multi-Component Inhomogeneous Bose-Einstein Condensates}

\author{Priyam Das}
   \email{daspriyam3@gmail.com}
   \affiliation{Institute of Nuclear Science, Hacettepe University, Ankara - 06800, Turkey}

\author{Ayan Khan}
   \affiliation{Gitam School of Technology, Gitam University, Bangalore - 562163, India}

\author{Prasanta K. Panigrahi}
   \affiliation{Indian Institute of Science Education and Research (IISER) Kolkata, Mohanpur - 741 246, India}

%\author{Priyam Das}
%   \ead{daspriyam3@gmail.com}
%   \address{Institute of Nuclear Science, Hacettepe University, Ankara - 06800, Turkey}
%
%\author{Ayan Khan}
%   \address{Gitam School of Technology, Gitam University, Bangalore, 562163, India}
%
%\author{Prasanta K. Panigrahi}
%   \address{Indian Institute of Science Education and Research (IISER) Kolkata, Mohanpur- 741 246, India}

\begin{abstract}
We study the dispersion mechanism of Lieb mode excitations of both single and multi component ultra-cold atomic Bose gas, subject to a harmonic confinement through chirp management. It is shown that in some parameter domain, the hole-like excitations lead to the soliton's negative mass regime, arising due to the coupling between chirp momentum and Kohn mode. In low momenta region the trap considerably affects the dispersion of the grey soliton, which opens a new window to observe Lieb-mode excitations. Further, we extend our analysis to binary condensate, which yields usual shape compatible grey-bright soliton pairs. The inter-species interaction induces a shift in the Lieb-mode excitations, where the pair can form a bound state. We emphasize that the present model provides an opportunity to study such excitations in the low momenta regime, as well as the formation of bound state in binary condensate.
\end{abstract}

%\PACS{
%      {67.85.Hj}{Bose-Einstein condensates}   \and
%      {03.75.Lm}{Solitons}                    \and
%      {03.75.Kk}{Dynamic properties}
%     }

%\pacs{67.85.Hj, 03.75.Mn, 03.75.Lm} \keywords{Bose-Einstein
%Condensate, Solitons, Lieb Mode} \maketitle

%\submitto{J. Phys. B: At. Mol. Opt. Phys.}

%%%%%%%%%%%%%%%%%%%%%%%%%%%%%%%%%%%%%%%%%%%%%%%%%%%%%%%%%%%%%%%%%%%%%%%%%%%%%%%%%%%%%%%%%
%                ONE COMPONENT BOSE-EINSTEIN CONDENSATE
%%%%%%%%%%%%%%%%%%%%%%%%%%%%%%%%%%%%%%%%%%%%%%%%%%%%%%%%%%%%%%%%%%%%%%%%%%%%%%%%%%%%%%%%%

{\let\newpage\relax\maketitle}

%\makeatletter
%\def\maketitle{%
%\par\textbf{\@title}%
%\par{\@author}%
%\par}
%\makeatother

\section{Introduction}

Since the first successful realization of the atomic Bose-Einstein condensate (BEC), the experiments, as well as theoretical works on this unique quantum state has advanced greatly by exploring different intriguing domains and expanding new horizons. By now we have obtained great controllability and tunability in the magneto-optic setup to cool, preserve and tune the atomic gases. This unique controllability has enabled multi facet research in ultra-cold atomic gases. Among them the quasi one dimensional (cigar shaped) condensate and associated soliton solutions of the mean-field  \GP equation, are quite well studied till date \cite{dalfovo,leggett,A.cornell,kett,sol,frantzeskakis,malomed}.

Nevertheless, the scope to study these non-linear systems are never ending \cite{gil,song,bishop} which has been
substantiated through several investigations during the last few years. Among them, we note the study of a trapped BEC, under the influence of an oscillating Gaussian potential, where both vortex pairs and solitons can be created by suitable amplitude modulations \cite{fujimoto}. In another work, splitting of the ground state of an attractively interacting BEC into two bright solitons with controlled relative phase and velocity has been explored \cite{billam}. The nonlinearity of Gross-Pitaevskii (GP) equation, describing BEC at the mean-field level, leads to these excitations \cite{carr,ldcarr,jackson,komineas}.

One may also recall that in 1963, Lieb discovered a collective excitation of the condensed bosons in a second quantized formulation, which exhibited a periodic dispersion, very different from the well-known Bogoliubov mode \cite{lieb}. The same was later identified with a complex grey soliton at the mean field level \cite{kulish}, which mimics a hole-like excitation of the condensate. Laboratory confirmation of this mode had been difficult due to the fact that, in the experimentally accessible low momenta regime, Lieb mode dispersion lies below that of the Bogoliubov excitations \cite{jackson,komineas,jack,pop,kom,ajack}. The instability of BEC for higher momentum values, where these two modes differ significantly, excluded this domain from observation, for quite some time.

Recent observation of oscillations between stable localized grey solitons in a cigar shaped BEC and vortex rings has led to significant interest in the Lieb mode, having complex envelope profile \cite{shomrani}. These solitons have been produced through collision of two BECs \cite{shomrani,weller}, when the collisional energy is reduced to a level, where it is comparable with the interaction energy. The wavelength of the laser, responsible for density modulation, is larger than the healing length. In this domain, nonlinearity plays a stronger role than dispersion and the interference effect leads to the formation of an array of grey solitons. At the lowest collisional energy, the interference pattern produces a single pair of grey solitons. These coherent structures may have relevance for atom interferometry \cite{sol,shomrani,weller,becker}. The presence of a harmonic trap and the time dependent coupling parameters, can non-trivially affect the soliton dynamics. In single component BEC, soliton and soliton trains in a trap with time varying scattering length and loss/gain have been analyzed \cite{atre}. A number of recent studies have explored the dark and bright solitons in the cigar-shaped BEC, revealing the nontrivial coupling between trap and soliton dynamics \cite{atre,ranjani1,ranjani2,utpal,rajendran,abhinav}. Hence, it is natural to inquire about the nature of the coupling between the trap and soliton geometry. The key question that arises here is  about the possibility of accessing the regime of Lieb-mode excitations, associated with a quantum grey soliton through the coupling between trap geometry and their excitations.

The aim of this letter is to address the above question and how the confining potential under certain circumstances affects the hole-like Lieb mode excitation. To start with, we show exact grey soliton solutions for one and two component BEC subject to a harmonic confinement. It is interesting to observe the dramatic collapse and revival of the atomic condensate by appropriate tailoring of the loss/gain term. Further investigations reveal that these chirped grey solitons can even be accelerated, compressed or brought to rest, when the effects of the harmonic trap on the width and amplitude of the solutions are explored. We then study the solitonic Lieb-mode dispersion and compare it with the usual sound (Bogoliubov) mode dispersion. It is interesting to observe that the temporal modulation of the chirping is found to be important to resolve the Lieb mode experimentally. The time modulated chirping enables us to find a small window where the hole-like excitation mode dominates over the sound mode before destabilizing the system. When the time dependency is overlooked, we extract the usual dispersion profile as shown in \cite{jackson}. One must note that, our formalism also elaborates on the control mechanism of chirping by means of external magnetic field. Thus the current formulation paves the way for their coherent control and manipulation at low energy, through the temporal modulation of the scattering length, as well as the trap \cite{atre,moores,longhi,kruglov,liang,engels,radha,alk}. The two component BECs (TBECs) show a rich variety of solutions (dark-dark, dark-bright and bright-bright), though in this paper, we restrict ourselves to the dark-bright pair. The effects of the inter-species interaction and harmonic trap on the dispersion of these hole-like excitations are investigated. In case of attractive interactions, the energy shows an upward shift with respect to the single component case. As we move from attractive regime towards repulsive, their energy decreases and exactly matches with the single component case, when inter-species interaction becomes zero. In the repulsive domain, the energy goes down further, entering into the negative regime, which leads to the formation of a bound state for these hole-like excitations. We argued that this bound state formation can manifest in the low energy mean-field theory.

The letter is organized in the following way, in Sec. 2 we present the exact analytical solutions for both one and two component case. These solutions lead to hole-like excitations or quantum grey soliton. We further show the dynamics in the presence of both regular and expulsive harmonic trap in Sec. 3, where these excitations undergo nonlinear compression in some appropriate parameter regime.  We investigate their dispersion relation in the later section (Sec. 4) and analyze the effects of the trap and interactions on the hole-like excitation associated with the Lieb mode. We draw our conclusion in Sec. 4 and discuss about the possible future goals.

\section{Quantum Grey Soliton}
\subsection*{Single Component}
The dynamics of BEC is well described at the mean-field level by the following 3D GP equation, with the order parameter $\Psi$:
% \cite{pethick}:
\begin{eqnarray}
i \hbar \frac{\partial \Psi}{\partial t} = - \frac{\hbar^{2}}{2 m} \vec{\nabla}^{2} \Psi + V({\bf r},t) \Psi +  U|\Psi|^{2} \Psi.
\end{eqnarray}
Here, $V({\bf r},t) = V(x,y) + V(z,t)$ and $U = 4 \pi \hbar^{2} a_{s}/m$ is the strength of the atom-atom interaction, with $a_{s}$ and $m$ being the scattering length and the mass of the atoms, respectively. At sufficiently low temperatures, the BEC confined in a strong transverse trap, $V(x,y)= \frac{1}{2}m \omega_{\perp}^{2} (x^{2}+ y^{2})$, with $\omega_{\perp}$ being the transverse frequency, can be made effectively one dimensional.

The appropriately scaled quasi-1D GP equation, in dimensionless units, can be written in the form \cite{atre,liang},
\begin{equation} \label{NLSE}
i \frac{\partial \psi}{\partial t} = -\frac{1}{2} \frac{\partial^{2} \psi}{\partial z^{2}} + \gamma(t)|\psi|^{2}\psi
+ \frac{1}{2} M(t) z^{2} \psi + i \frac{g(t)}{2} \psi - \frac{\nu(t)}{2}\psi.
\nonumber\\
\end{equation}
Here, the strength of the atom-atom interaction and the spring
constant are $\gamma(t)=2a_{s}(t)/a_B$ and
$M(t)=\omega_{0}^{2}(t)/\omega_{\perp}^{2}$, respectively, with $a_B$
being the Bohr radius. For the sake of generality, we have kept $M(t)$
time dependent. Regular and expulsive oscillator potentials correspond
to $M > 0$ and $M < 0$, respectively.

In order to solve the above nonlinear equation, a self-similar method is adopted here, which is straightforward and as a result of it, one can clearly see the dependence of the trapping potential and the time dependent nonlinearities. This method has already been elaborated in detail by Atre {\it et. al.} \cite{atre}. With this approach, we assume the following ansatz solution:
\begin{eqnarray}
\psi(z,t) = \sqrt{A(t) \sigma(\xi)} e^{i \left[\Phi(z,t) + \chi(\xi) + \frac{G(t)}{2} \right]},
\end{eqnarray}
where, we described the solutions in terms of the center of mass co-ordinate $\xi(z,t) = A(t) (z - l(t))$ and $G(t) = \int_{0}^{t} g(t')dt'$ represents the phenomenological loss/gain term.
The kinematic phase has a quadratic form: $\Phi(z,t) = a(t) + b(t) z - \frac{1}{2} c(t)z^{2}$, exhibiting chirping, where
$a(t)=a_{0}-\frac{1-\bar{\mu}}{2}\int_{0}^{t}A^{2}(t')dt'$  and $b(t) =
A(t)$. Here $\bar{\mu}=-2\kappa\sigma_{0}=\mu+\lambda$;
$\mu = 2 \nu(t)/A^{2}(t)$ is the chemical potential in the moving co-ordinate frame and $\lambda$, respectively, are the chemical potential and a constant parameter controlling the energy of the excitation. The coefficient of chirping $c(t)$ can be determined from the following Riccati type equation, $\frac{\partial c(t)}{\partial t} - c^{2}(t) = M(t)$. It is worth pointing out that the above Riccati equation can be expressed as a Schr\"odinger eigenvalue problem by incorporating the appropriate transformation as shown in \cite{atre}.

The width of the soliton profile is given by,  $A(t) = A_{0} e^{\int_{0}^{t} c(t')dt'}$. It is worth noting that center of mass motion (COM) has a direct connection with the Kohn theorem \cite{kohn}. For better understanding, we separate the governing equation of COM: $\frac{\partial l(t)}{\partial t} + c(t) l(t) = A(t) u$. The Kohn theorem states that the center of mass motion oscillates exactly with the trap frequency, undergoing sinusoidal oscillations. When the trap frequency varies with time, the COM motion ceases to decouple from the trap, leading to a nontrivial behavior as discussed in detail by the authors in \cite{das2015}. In order to gain complete control of the spatio-temporal dynamics of the condensates and these hole-like excitations, we separate the resulting equation of the COM of these excitations,
\begin{eqnarray}
\frac{\partial^{2} l(t)}{\partial t^{2}} + M(t)  l(t) = 0.
\end{eqnarray}

\begin{figure}[t]
\begin{center}
\includegraphics[scale=0.4]{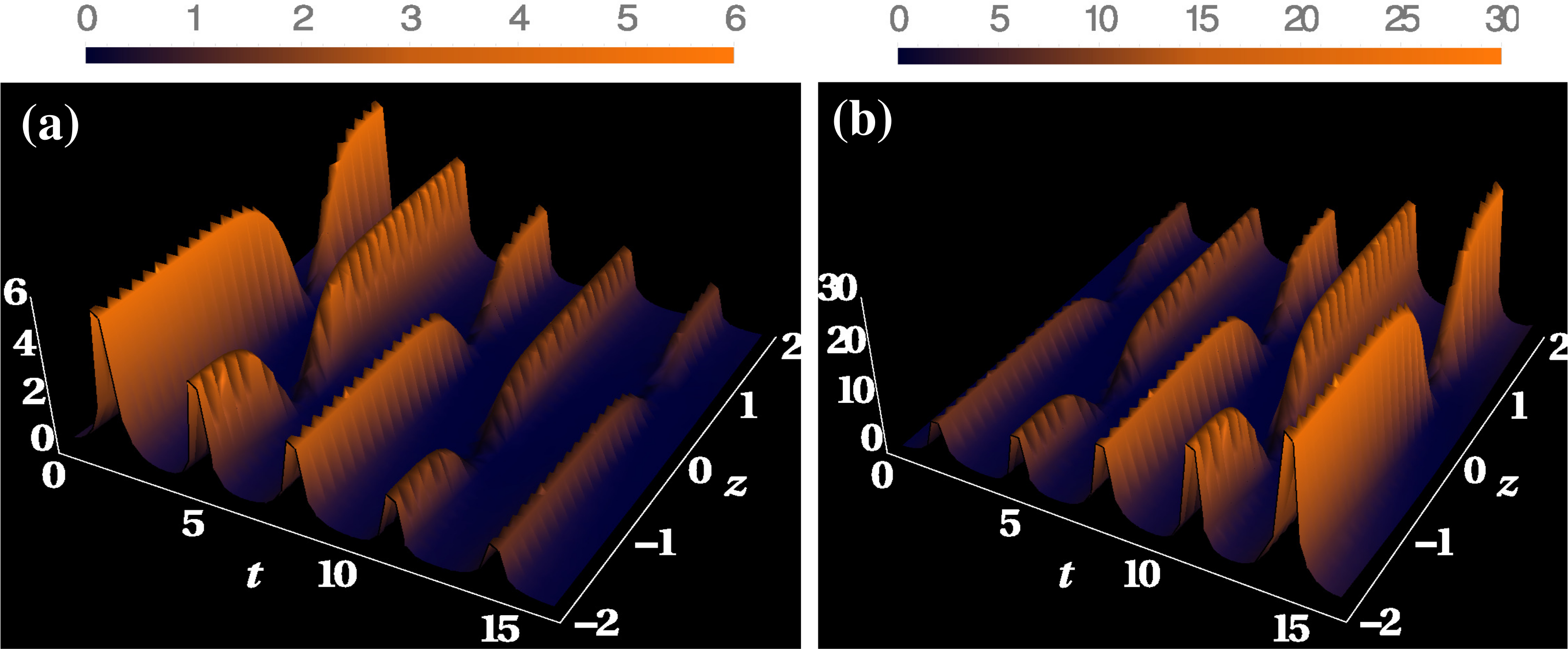}
\caption{ Collapse and revival of the atomic condensate.  (a) depicts the collapse and revival with increasing amplitude by appropriate tailoring the loss/gain ($\beta = 0.1$).  (b) represents the same collapse and revival of the atomic condensate with decreasing amplitude for a different value of loss/gain ($\beta = -0.1$). In both the cases, the grey soliton shows small oscillation in space due to the presence of the harmonic confinement. The other parameter values are as follows: $A_{0} = 1$, $a_{0} = 1.5$, $l_{0} = 0.7$ and $\lambda = 0.9$, $\gamma_{0} = 0.5$ and $M(t) = 1$.}
\label{collapse}
\end{center}
\end{figure}

From current conservation, amounting to solving the imaginary part of the GP equation, one obtains, $\frac{\partial\chi(\xi)}{\partial \xi} = u (1 - \frac{\sigma_{0}}{\sigma(\xi)})$, where, $\sigma_{0}$ is the Thomas-Fermi background density of the atoms in the center of mass frame $\xi(z,t)$. In this frame, the density equation can be cast in the convenient form \cite{jackson},
\begin{eqnarray}\label{hydro2}
\Bigg(\frac{\partial \sqrt{\sigma(\xi)}}{\partial \xi}\Bigg)^{2}&=&(\kappa \sigma -u^{2}) \frac{(\sigma - \sigma_{0})^{2}}{2\sigma} \textrm{,}
\end{eqnarray}
where,  $\gamma(t) = \gamma_{0} e^{-G} A(t)/A_{0}$ with $\kappa = \frac{\gamma_{0}}{A_{0}}$. The solution of Eq. (\ref{hydro2}) takes the self-similar form,
\begin{eqnarray}\label{solution}
\sigma(\xi) = \sigma_{0} - \sigma_{0} \cos^{2}{\theta} \,\, \textrm{sech}^{2} \, \left[\frac{\cos{\theta}}{\zeta}\xi(z,t)  \right],
\label{sol_psi}
\end{eqnarray}
where the Mach angle is given by, $\theta = \sin^{-1} \frac{V}{C_{s}} = \sin^{-1} \frac{u}{c_{s}}$. Here, $V = A(t) u$ is the velocity of the solitons, which is bounded by the sound velocity $C_{s} = A(t) \sqrt{\kappa \sigma_{0}} = A(t) c_{s}$; $c_{s}$ is the sound velocity in the absence of the harmonic trap. By appropriate tailoring of the loss/gain ($g(t) = - \beta t$) leads to an interesting phenomena: collapse and revival of the atomic condensate. In Fig.(\ref{collapse}), the collapse and revival with decreasing (\ref{collapse}a) and increasing (\ref{collapse}b) amplitudes of the quantum grey soliton are shown for positive and negative loss/gain, $\beta=0.1$ and $\beta=-0.1$, respectively. The collapse and revival of this hole-like excitations are connected with the interplay between trap geometry and time modulated interactions, which can be tailored through the loss/gain term as well. In the limiting case, i.e., when the harmonic trap is switched off, all the expressions match with the known results \cite{jackson,komineas}.

\subsection*{Binary condensate}

Binary mixtures of BECs are available in experimental studies. Most of the binary mixtures contain two different hyperfine states of the same atomic species, such as $^{87}Rb$ \cite{cornell}, and $^{23}Na$ \cite{andrews}. A BEC has also been created in a hetero-nuclear mixture of $^{41}K$ and $^{87}Rb$ \cite{simoni}. The two component BECs (TBECs), present a novel and fundamentally different scenario for their ground state structure, as compared to the single component case. Theoretically, TBECs have been widely studied and various structures have been identified, such as dark-dark \cite{ohberg} and dark-bright solitons \cite{busch}. Very recently, Hamner {\it et.al.}, presented the first experimental observation of dark-bright soliton trains in superfluid-superfluid counterflow, and showed that this induced a modulational instability in this miscible system \cite{hamner}. In another paper, Shimodaira {\it et. al.}, considered  two component BEC rotating in a toroidal trap and showed the transition from immiscible to miscible condensate \cite{shimodaira}.

In this analysis, we consider the binary condensate, placed in two different hyperfine states of $^{87}Rb$ atoms. The TBEC are described by the coupled GP equations, which can easily be put into dimensionless form, similar to the single component:
\begin{eqnarray}
i \frac{\partial \psi_{j}}{\partial t} = - \frac{1}{2} \frac{\partial^{2} \psi_{j}}{\partial z^{2}} + \frac{1}{2} M(t) z^{2} \psi_{j} &+& \sum_{k=a,b} \gamma_{j k}(t) |\psi_{k}|^{2}\psi_{j} \nonumber \\  &-& \nu_{j}(t) \psi_{j} + i \frac{g_{j}(t)}{2} \psi_{j}
\label{coup.nlse2},
\end{eqnarray}
where, $\psi_{j}(z,t)$ with  $j = a, b$  refers the wave function of the first and second component, respectively. The intra-species interactions between the atoms are given by, $\gamma_{j j}(t) = 2 \frac{a_{j j}(t)}{a_{B}}$, which in fiber optics language is known as self-phase modulation, while the inter-species interactions are denoted as, $\gamma_{ab}(t) = 2 \frac{a_{ab}(t)}{a_{B}}$, known as cross-phase modulation. Without any loss of generality, we here assumed $\gamma_{ab}(t) = \gamma_{ba}(t)$ and from now onwards for notational convenience, we denote $\gamma_{j j}(t) = \gamma_{j}(t)$. $\nu_{j}(t) = \mu_{j} A^{2}(t)$ the chemical	 potential of $j^{th}$ component. The BEC wave functions are taken in the form,
\begin{eqnarray}
\psi_{j}(z,t) &=& \sqrt{A(t) \sigma_{j}(\xi)}  e^{i \chi_{j}(T)- i \frac{1}{2} c(t) z^{2} + i \frac{G_{j}(t)}{2}},
\label{ansatz}
\end{eqnarray}
where, the parameters $\xi(z,t)$, $A(t)$ and $c(t)$ are same as in the
previous case, with $l(t) = \int_{0}^{t} v(t') dt'$. In this case, $G_{j}(t) = \int^{t}_{0} g_{j}(t') dt'$ is the phenomenological loss/gain term for first ($j=a$) and second component ($j=b$). For the sake
of simplicity, we have considered here: $a (t) = b(t) = 0$. Therefore, from current
conservation, we obtain the following two hydrodynamic equations,
\begin{eqnarray}
\frac{\partial \chi_{a}}{\partial \xi} = u (1 - \frac{\sigma_{0}}{\sigma_{a}}) \textrm{\,\,\,and \,\,\,}
\frac{\partial \chi_{b}}{\partial \xi} = u.
\end{eqnarray}

\begin{figure}[t]
\begin{center}
\includegraphics[scale = 0.35]{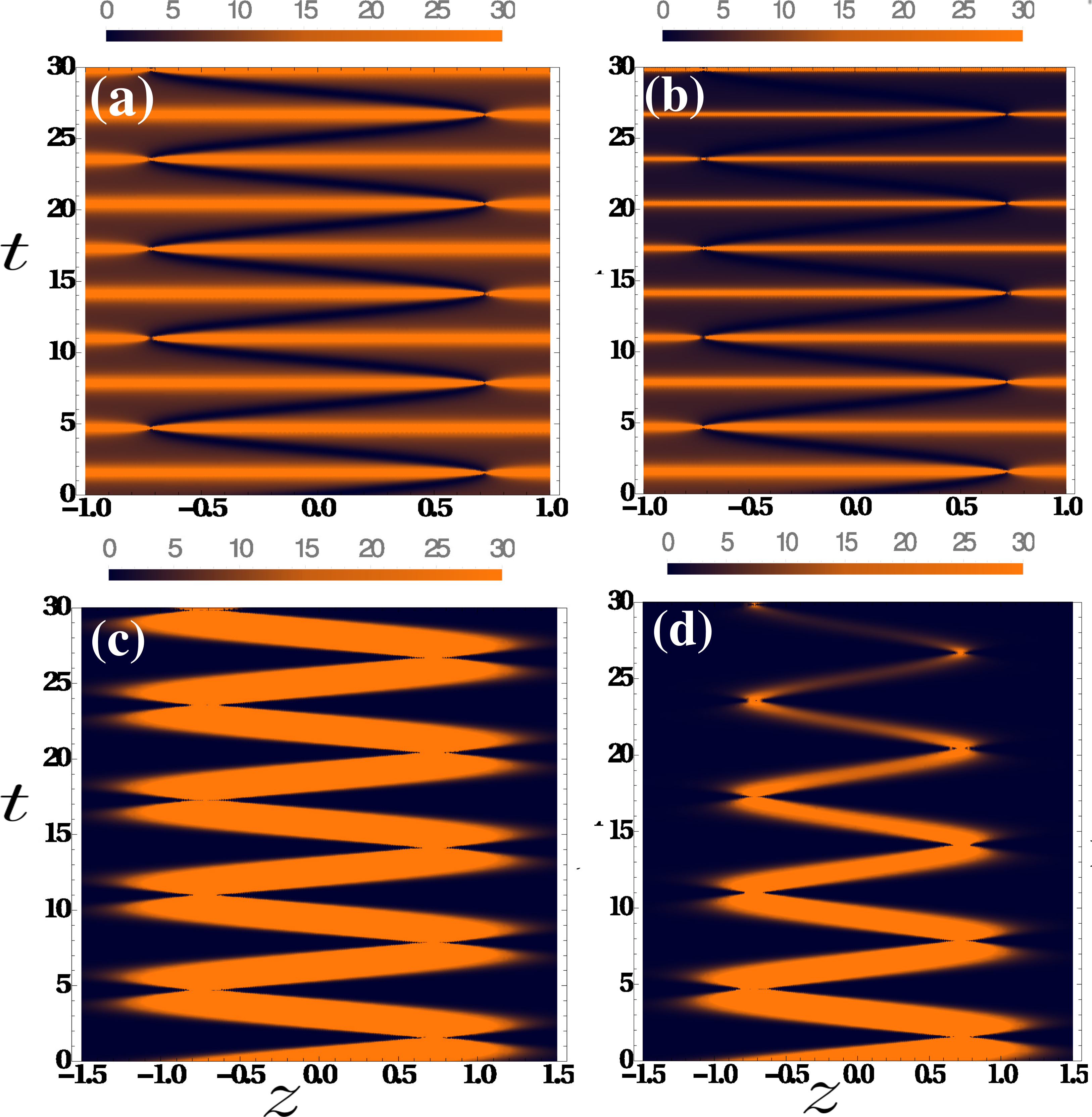}
\caption{The figures show the variation of the soliton profile for $\theta = \pi/16$. Fig. (a) and (b),
respectively show the density profile of the hole-like excitations for $\beta_{a} = 0$ and $\beta_{a} = 0.05$. The bright soliton profile is shown in (c) and (d) for $\beta_{b} = 0$ and $\beta_{b} = 0.2$, respectively. We consider here, $\kappa_a = 4$, $\kappa_b = 0.5$ and $\kappa_a = 2.4$. The other parameter values are same as in Fig. (\ref{collapse}).}
\label{soliton}
\end{center}
\end{figure}

The phase of the second component is taken to be independent of its density, implying that the
corresponding superfluid velocity is constant.  The center of mass
motion of the soliton in this case, turns out to be: $l(t) = l_{0} e^{- \int^{t}_{0} c(t') dt'}$,
which can be controlled by modulating the trapping frequency. The real
part of the coupled equations leads to the following soliton solutions of the
first and second condensates respectively,
\begin{eqnarray}
\sigma_{a}(\xi) &=& \sigma_{0}(1 - \cos^{2}\theta \textrm{sech}^{2}\Big(\frac{\cos\theta}{\zeta} \xi(z,t) \Big) \label{dark},\\ %\textrm{with\,\,\,}
%\sigma_{b}(\xi) &=& \sigma_{0} \cos^{2}\theta \frac{\kappa_{a} -  \kappa_{ab}}{\kappa_{ab} - \kappa_{b}} \textrm{sech}^{2}\Big(\frac{ \cos\theta}{\zeta} \xi(z,t) \Big)
\sigma_{b}(\xi) &=&  \frac{N_{b} \cos\theta}{2 \zeta} \textrm{sech}^{2}\Big(\frac{ \cos\theta}{\zeta} \xi(z,t) \Big),
\label{bright}
\end{eqnarray}
where, $N_{b} = \int dz |\psi_{b}(z,t)|^{2}$ is the rescaled number of atoms in state $b$. The interaction parameters are scaled as : $\kappa_{j} = \frac{\gamma_{j}(t)}{A(t)}$, with $j=a, b, ab$. Akin to the single component case, this shows that the frequency of the harmonic oscillator is related to the interaction parameter through the chirping term. The chirped phase in this case plays a very crucial role in determining the various parameters associated with solitons.  The healing length is found to be of the form: $\zeta = \sqrt{(\kappa_{ab} - \kappa_{b})/[\sigma_{0} (\kappa^{2}_{ab} - \kappa_{a} \kappa_{b})}]$, and $\theta = \sin^{-1} \frac{u}{c_{s}}$. The soliton's velocity $u$ achieves a maximum value $c_{s} = \frac{1}{\zeta}$, when $\theta = \frac{\pi}{2}$. This corresponds to a constant solution $\sigma_{a} = \sigma_{0}$.  The $\theta = 0$ limit leads to a pure dark soliton solution $\sigma_{a} = \sigma_{0} \tanh[\xi(z,t)/\zeta]$, which is static, since $u = 0$.  The chemical potential of the first component is similar to the single component case, whereas, for second component it strongly depends on the inter-species interaction and healing length :  $\mu_{b} = \kappa_{ab} \sigma_{0} - \frac{1}{2}u^{2} - \frac{\cos^{2}\theta}{2 \zeta^{2}}$. Similar to the single component case, we assume here the loss/gain term as $g_{j}(t) = \beta_{j} t$.

Fig. (\ref{soliton}) depicts the corresponding density profile of the grey and bright solitons. Fig. (\ref{soliton}a) shows the propagation of a grey soliton, which oscillates with time due to the presence of the harmonic trap, without the presence of loss/gain ($\beta = 0$). The phase associated with this excitation changes at each oscillating point. Fig. (\ref{soliton}b) refers to the density distribution of the hole-like excitation with loss/gain ($\beta = 0.05$). The amplitude of the atomic distribution diminishes with time. The density profile of the bright soliton, characterized by a localized maximum is shown in  Fig. (\ref{soliton}c). It is clear from the figure that bright soliton also oscillates at the same frequency as in the case of grey soliton. This frequency of oscillation is directly related to the frequency of the harmonic trap. Fig. (\ref{soliton}d) depicts the same in the presence of loss/gain ($\beta = 0.2$). The amplitude of the bright soliton decreases with time as well, though the frequency of the oscillations remains same. One can vary the soliton profile by changing the nonlinearities and the trap frequency. In the real experiment, the nonlinearities and the confining harmonic trap can be changed by tuning the magnetic field induced Feshbach resonances and optical trap respectively.

\section{Nonlinear compression of quantum grey soliton}

\subsection*{Single component}

\begin{figure}[h]
\begin{center}
\includegraphics[height=3.75cm,width=8.5cm]{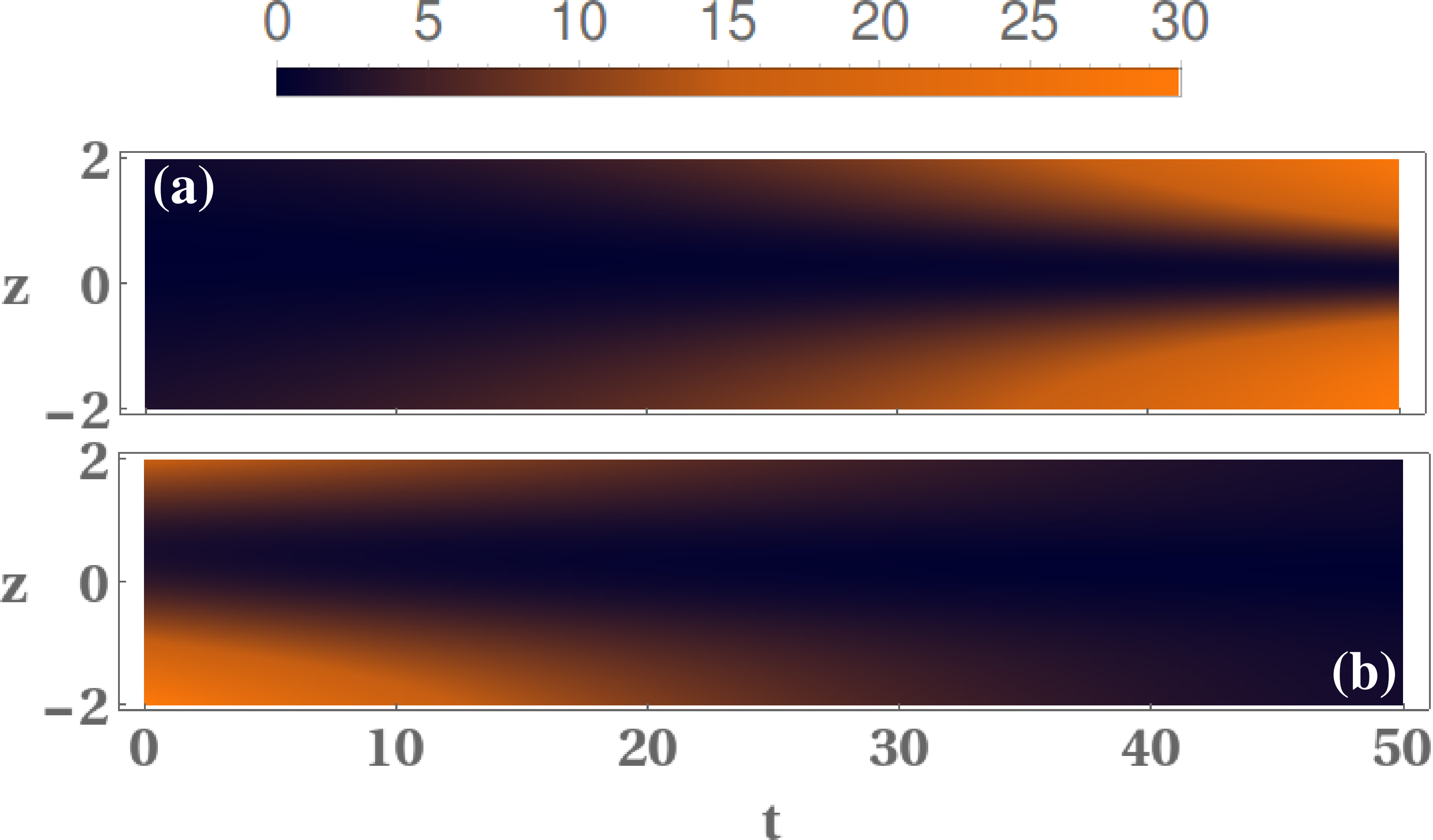}
\caption{(a) The nonlinear compressin of the quantum grey soliton ($\theta = \pi/16$) as it propagates. The width of the soliton decreases with the increasing depth. (b) The hole-like excitations spread out in presence of the loss/gain term ($\beta = 0.12$). The other parameter values are same as in Fig. (\ref{collapse}). }
\label{compression1D}
\end{center}
\end{figure}

Unlike adiabatic soliton compression, the present system takes an advantage of the exact solution to the nonlinear
Schr\"odinger equation for chirped grey soliton evolution. In order to see the nonlinear compression, as well as for
experimental realization, we present the following example. For simplicity, we consider the loss/gain term $G = 0$. The
Feshbach managed scattering length can be tuned as: $a_{s}(t) = a_{0} e^{\lambda t}$ \cite{liang}. In the presence of an
expulsive parabolic oscillator with $M(t) = - \lambda^{2}$, we find the chirping coefficient as: $c(t) = \lambda$.
Consecutively, all other parameters are obtained: $A(t) = A_{0} e^{\lambda t}$ and $l(t) = l_{0} e^{- \lambda t}$. These solitons are necessarily chirped and can even be accelerated, compressed or brought to rest. Fig. (\ref{compression1D}a) depicts the nonlinear compression of the dark soliton; one notices that with the increasing value of the scattering length, the dark soliton has an increase in amplitude and a compression in its width. The soliton can be brought to rest at $\theta = 0$. As we increase the value of $\theta$, the velocity of the soliton increases and the soliton propagates along the longitudinal direction.

Therefore, the acceleration and the compression of the soliton can be controlled through trap modulation and tuning of the scattering length by means of Feshbach resonances. In the following section, we extend our analysis to the two component case, in order to see the effect of intra- and inter-atomic interactions, on grey soliton dynamics.

\subsection*{Binary BEC}

\begin{figure}[t]
\begin{center}
    \includegraphics[scale = 0.35]{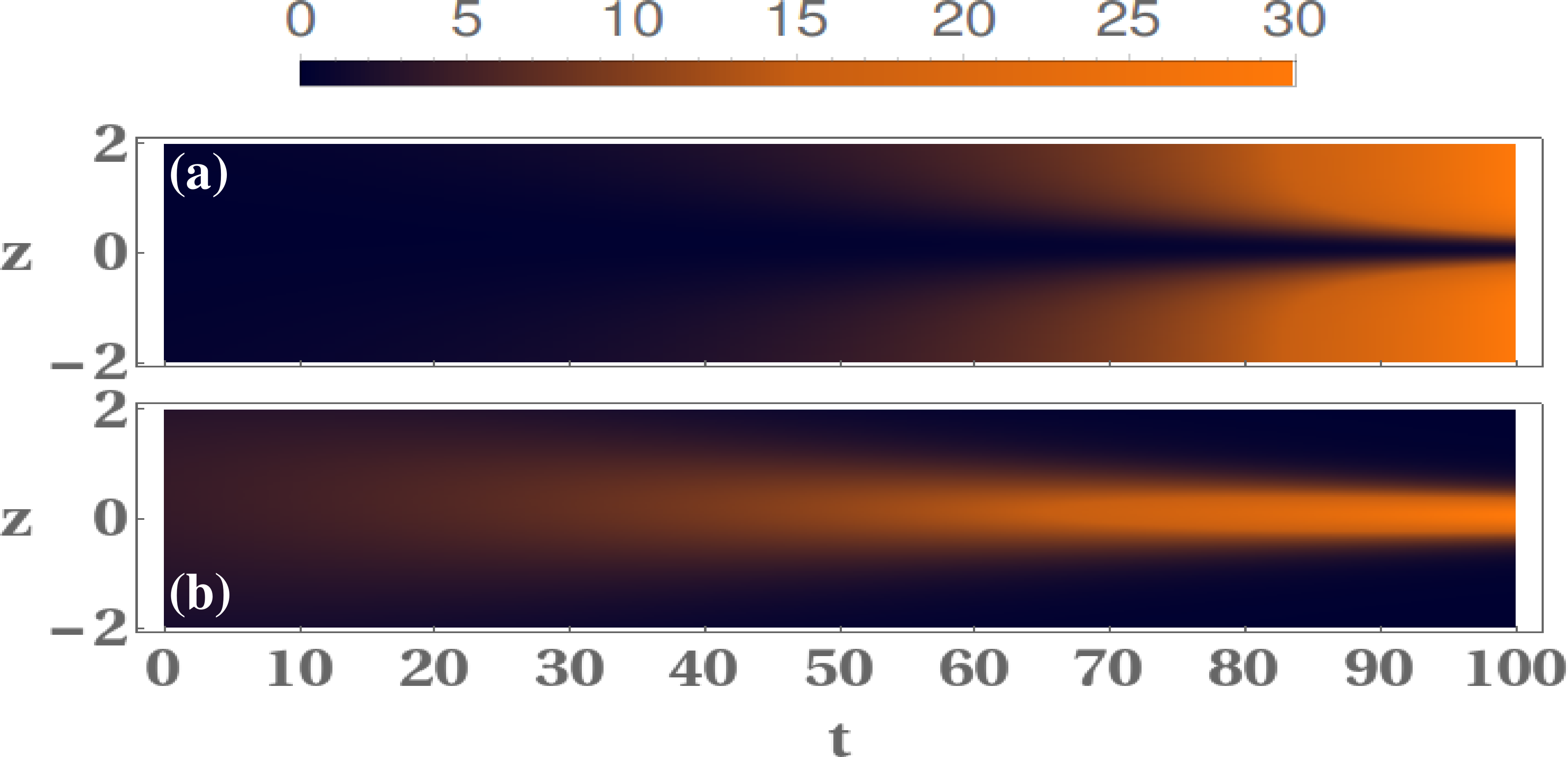}
\caption{The nonlinear compression of the dark and bright soliton is shown for $\theta = \pi/16$. As is seen, the width of the dark-bright solitons decreases with increasing amplitude. The other parameter values are same as in Fig. (\ref{collapse}) and (\ref{soliton}).}
\label{compression2bec}
\end{center}
\end{figure}

There are many tools, which are being used to control and manipulate the various parameters of solitons and induce changes in their shapes, which is very useful for developing many applications of BECs. One way is to vary the atomic scattering length by means of external magnetic field, known as Feshbach resonance. We here demonstrate that in the presence of an "expulsive" trap, the variation of scattering length in a particular manner leads to a nonlinear compression of both dark and bright soliton simultaneously. Fig. \ref{compression2bec}(a) shows tha compression of the width of a dark soliton, as seen in case of single component. The nonlinear compression of the bright soliton is shown in Fig. \ref{compression2bec}(b), which shows that the width of the bright soliton can also be compressed. As the scattering length increases, the amplitude of the bright soliton increases. At $\theta = 0$, the dark soliton becomes a black soliton with its velocity $u = 0$. The amplitude of the bright soliton becomes maximum at this point. As we increase $\theta$, the amplitude of the bright soliton decreases and vanishes at $\theta = \pi/2$, whereas, the dark soliton takes a constant value. At this point, the velocity of the solitons achieves its maximum value, i.e., the sound velocity. This technique can be used to modulate the bright soliton into very high local density in an inverted harmonic oscillator potential.

The above description for solutions of multi-component BEC now provides us the platform to investigate different modes associated with this system. In the following section we investigate the dispersion relations of Lieb and Bogoliubov modes carefully and identify a domain, where Lieb mode becomes experimentally realizable in the background of sound mode.

\section{Lieb-mode excitations}

\subsection*{Single component}

To study the regime of Lieb-mode, one requires the calculation of canonical energy and momenta of the system.
The energy of the grey soliton can be evaluated by subtracting the energy due to the contribution of the background term, such that $E = W - W_{0}$, where,
%\begin{widetext}
\begin{eqnarray}
W &=& \int\left[\frac{1}{2} \left(\frac{\partial \psi^{*}}{\partial z}
  \frac{\partial \psi}{\partial z}\right) + \frac{1}{2}
  \gamma(t)(\psi^{*}\psi)^{2}) +
  \frac{1}{2}M(t)z^{2}\psi^{*}\psi\right]dz. \nonumber \\ \label{energy.eq}
\end{eqnarray}
The energy expression $W_{0}$ corresponds to the profile
$\psi_{0}=\sqrt{A(t) \sigma_{0}}e^{i\Phi(z,t)+G/2}$; it has a kinematic phase in a
trap of similar parameter values. Given the solitary wave profile of
Eq. (\ref{sol_psi}), Eq. (\ref{energy.eq}) yields,
\begin{eqnarray}
E &=& e^{G} \Big[\frac{4}{3}\kappa A^{2}\sigma_{0}^{2} \zeta
  \cos^{3}\theta - (c^{2}+M) \Bigg(\frac{\zeta^{2}\sigma_{0}}{A^{2}
    \cos\theta} \frac{\pi^{2}}{12} \nonumber\\ & & + l^{2} \zeta \sigma_{0}
  \cos\theta\Bigg)  + 2 A b u
  \sigma_{0}\zeta\cos\theta+b^2\sigma_{0}\zeta\cos\theta\Big]\label{energy},
\end{eqnarray}
where, the exponential pre-factor $e^G$ shows the effect due to the loss/gain of atoms from the condensate. The first term in the expression represents the contribution from the solitonic energy. The term thereafter accounts for the oscillator and chirp contributions; it is singular at $\theta=\pi/2$. The third term, linear in $b(t)$, arises from the coupling of BEC momentum with the phase of the soliton.The last term being quadratic in $b(t)$, represents BEC translational energy.
One must recall that we have started with a very general structure, keeping several equation parameters as time dependent. If one looks at the Riccati equation and considers $c(t)$ as independent of time, then we obtain a condition: $c^2+M=0$. This removes the energy divergence in Eq.(\ref{energy}) and it exactly matches with \cite{jackson,komineas}. This situation we have already noted in our preceding section and showed that the subsequent expressions of the equation parameters are in accordance with literature \cite{liang}. Thus, through a controlled modulation of the external magnetic field, we are actually able to take into account the external trap and extract the usual result through a complete analytical model. However, the temporal dependence of the chirping also carries some significant information, which we will elaborate after presenting the canonical momentum calculation.

\begin{figure}[t]
\begin{center}
%\subfigure[$ $]{
  \includegraphics[scale=0.37]{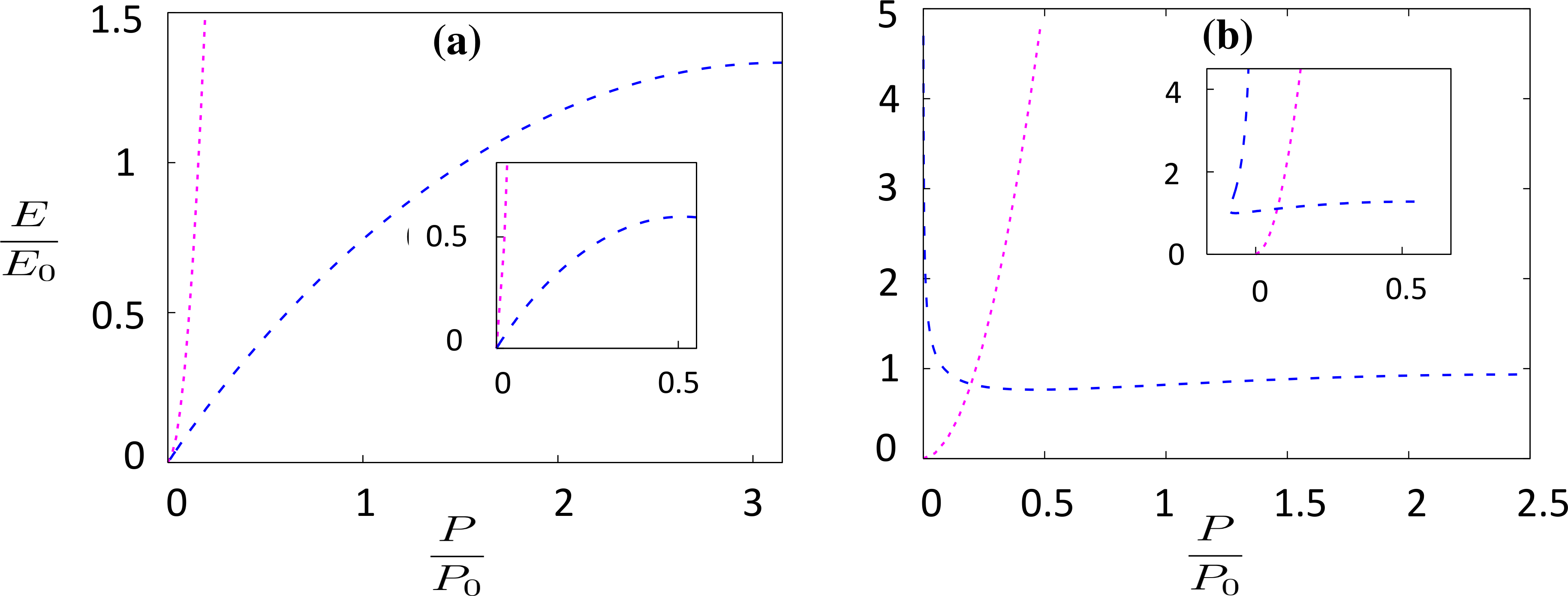}
%\subfigure[$ $]{
%  \includegraphics[scale=0.3]{lieb_bogo_multi_mod.eps}}
\caption{Dispersion relations for the soliton (blue dashed line) and Bogoliubov mode (pink dotted line). (a) The behavior of Lieb and Bogoliubov modes are shown at $t=0$. Inset reveals that the periodicity differs when $t\neq0$. (b) The Lieb mode dispersion for $b=0$, in presence of harmonic trap. Inset depicts the mode behavior, when $b\neq0$. The other parameter values are same as in Fig. (\ref{collapse}).}
\label{lieb_bogo}
\end{center}
\end{figure}

The canonical momentum of the solitary wave is given by,
\begin{eqnarray}
P &=& - i \int{\psi^{*}\frac{\partial \psi}{\partial z}}dz \nonumber\\
  &=& e^{G} \left[C_{s} \zeta \sigma_{0} \left(\pi\frac{u}{|u|} - \sin 2\theta -2 \theta\right) - 2 b \zeta \sigma_{0}
  \cos\theta\right] \label{momentum}
\end{eqnarray}

The first term can be attributed to soliton, while the second term arises from the chirped momentum. The energy and momentum are, respectively, normalized by $E_{0}=\kappa \zeta (A_{0}\sigma_{0})^2$ and $P_{0} = A_{0}c_{s}\sigma_{0}\zeta$. It is worth mentioning that the soliton velocity can also be computed from the hydrodynamic relation: $\frac{\partial E}{\partial P} = A u$, which matches with the earlier obtained results. In the limiting case, i.e., when the trap is switched off, all the expressions match with the known results \cite{jackson,komineas}. Eq. (\ref{momentum}) implies that a maximum momentum of $P_{max} = e^{G} (C_{s} \zeta \sigma_{0} \pi - 2 b \zeta \sigma_{0})$ is obtained for $u = 0$. The Lieb mode terminates at $P = P_{max}$. The Bogoliubov or sound mode is calculated using the usual prescription, by the application of a perturbation and subsequent linearization of the perturbed \GP equation.

\begin{figure}[t]
\begin{center}
%\subfigure[$ $]{
  \includegraphics[scale=0.4]{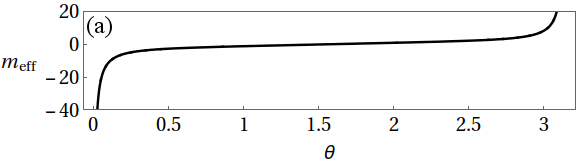}
%\subfigure[$ $]{
  \includegraphics[scale=0.4]{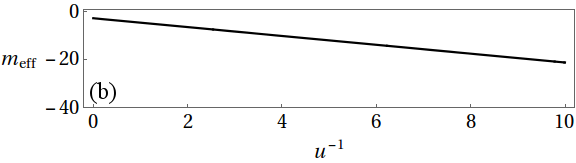}
\caption{The effective mass profile is shown as a function of (a) Mach angle $\theta$ and (b) velocity $u$. This is akin to the scenario of negative mass in the low momenta regimes. (b) shows the emergence of negative mass regime for small values of $u$ and the negative slope gives rise to the momentum. The parameter values are same as in Fig. (\ref{collapse}).}
\label{mass}
\end{center}
\end{figure}

We present different consequences of dispersion in Fig.(\ref{lieb_bogo}). In Fig. (\ref{lieb_bogo}a) the dispersion associated with the grey soliton profile with well known $2 \pi$ periodicity with $t=0$ and without the trap. However, when $t\neq0$ in the absence of trap, the characteristic periodicity of the solitonic energy will change. We have shown this behavior in the inset of Fig.(\ref{lieb_bogo}a). Now we  would like to draw the attention to the Fig.(\ref{lieb_bogo}b). In Fig.(\ref{lieb_bogo}b) the solitonic energy shows a divergence. One may try to avoid this situation in experiment, we show here that this can offer new possibilities as well. We can clearly see that in a small momentum window, the solitonic energy is much higher than the sound mode, implying that it is possible to observe this solitonic mode in a restricted region with controlled chirping (which can be controlled experimentally) before the system gets destabilized. Further, the presence of the translational motion of BEC considerably affects the grey soliton and one can observe that in a certain low momenta region, the soliton's momentum can become negative, as depicted in the inset of Fig. (\ref{lieb_bogo}b), thereby suggesting a region of negative soliton mass. More precisely, taking into account the momentum conservation, the time-dependent momentum term in chirping $b(t)$ is expected to get strongly correlated with the COM of the soliton. This strong coupling between the chirped momentum and COM enables us to achieve the regime of negative mass. The strong coupling between the chirp momentum and COM can be understood from the effective mass profile. Therefore, we obtain the effective mass, as shown in \cite{kevre2015,malomed2004},
\begin{eqnarray}
m_{eff} = P/V = m_{0} \Xi(\theta, u),
\end{eqnarray}
where, $m_{0} = A_{0} \sigma_{0} \zeta$ is a scaling factor. The factor $\Xi(\theta, u)$ can easily be obtained using Eq. (\ref{momentum}). The effective mass profile is shown in Fig. (\ref{mass}). For positive $u$, current conservation yields, $\frac{\partial \chi(\xi)}{\partial \xi}$ to be negative for hole-like excitations. This accumulates a net phase difference (negative) as the grey soliton starts propagating. The combine effects of this negative phase accumulation and chirped momentum enable us to enter into the negative mass regime.  The variation of the effective mass as a function of $\theta$, as shown in Fig. (\ref{mass}a) provides a better understanding. It shows that small values of $\theta$, resembles the small momentum values, the effective mass becomes negative. This is akin to the scenario depicted in the inset of Fig. (\ref{lieb_bogo}b). Fig. (\ref{mass}b) replicates the same as a function of the inverse velocity ($u^{-1}$). We observe that the effective mass becomes negative for small values of $u$ and the slope (negative) gives rise to the momentum. Hence, it is possible to achieve the negative mass regime, which emerges in a small momentum window. We emphasize that this regime is possible to achieve experimentally though the chirp management within the small momentum window before the system gets destabilized. In this context, it is worth mentioning that negative mass regime for gap solitons has already been proposed \cite{boris} and observed experimentally \cite{anker}.

As a next step we show that similar dispersion mechanism also arises in the two component case, which reveals much richer physics, due to the presence of the inter-species interaction.
%, where the energy achieves its maximum value.

\subsection*{Formation of bound state}

The energy of the grey soliton in case of TBECs can be obtained from Eq. (\ref{energy.eq}),
%\begin{eqnarray}
%W &=& \int \Big[\frac{1}{2}\frac{\partial \psi^{*}_{a}}{\partial z}  \frac{\partial \psi_{a}}{\partial z} + \frac{1}{2} M(t)z^{2}  (\psi^{*}_{a} \psi_{a})  + \frac{1}{2}\gamma_{a}(t)
 % (\psi^{*}_{a} \psi_{a})^{2} \nonumber \\ & & + \gamma_{ab}(t) (\psi^{*}_{a} \psi_{a})  (\psi^{*}_{b}\psi_{b}) - \mu  (\psi^{*}_{a}\psi_{a})\Big]. \label{energy_def}
%\end{eqnarray}
where, the background energy $W_{0}$ is obtained by assuming a constant solution. Unlike the single component case, we consider here $b(t)=0$. Since, we are interested in the Lieb-mode excitation, we concentrate only to the energy of the quantum grey soliton. The parameter values, used for the constant solution are same as the previous case. Using the solitary (grey) density profiles in Eq. (\ref{dark}) and (\ref{bright}), Eq. (\ref{energy.eq}) yields,
\begin{eqnarray}
E_{a} &=& \frac{2}{3 \zeta} \sigma_{0} A^{2}(t)
\cos^{3}\theta + \zeta \sigma_{0} (c^{2}(t) + M^{2}(t)/2)
\Bigg(\frac{\pi^{2} \zeta^{3} \sigma_{0}}{6 A^{2} \cos \theta} \nonumber \\ & & + 2
l^{2}(t) \zeta \sigma_{0} \cos \theta \Bigg) +
\frac{2}{3} \kappa_{a} \sigma^{2}_{0} A^{2} \cos^{3}\theta \nonumber \\ & & -
\frac{4}{3} \zeta \kappa_{ab}\sigma_{0} b^{2} A^{2}(t) \cos \theta.
\end{eqnarray}

\begin{figure}[h!]
\begin{center}
\includegraphics[scale = 0.3]{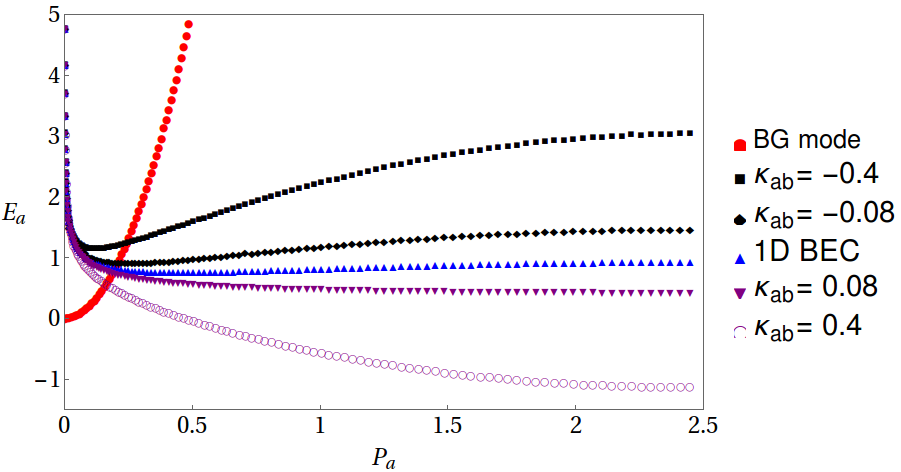}
\caption{The figure shows the dispersion relation of the grey soliton for different values of inter-species interactions. As shown, for attractive (repulsive) interaction the dispersion lies above (below) the dispersion of the single component one. The other parameter values are same as in Fig. (\ref{collapse}) and (\ref{soliton}).}
\label{lm-tbec}
\end{center}
\end{figure}

%$\psi_{a0} = \sqrt{A(t) \sigma_{0}} e^{-i c(t)x^{2}/2}$
Similar to the single component case, the contributions from the
harmonic trap and chirping to the energy turns out to be singular at
$\pi/2$. The third and fourth terms are due to the presence of intra- and inter-species ineractions,
respectively. The canonical momentum of the grey soliton is found
to be,
\begin{equation}
P_{a} = \zeta \sigma_{0} A(t) u_{s} (\pi \frac{u}{|u|} - 2 \theta - \sin 2\theta) + 2 \zeta \sigma_{0} c(t) l(t) \cos \theta.
\end{equation}
The first term in the momentum expression is the momentum of the soliton and the second term is due to the presence of chirping. The effect of the inter-species interaction on dispersion profile is investigated and is shown in Fig.(\ref{lm-tbec}). At low momenta, the energy is found to be very high. The fact that the presence of trap considerably affects the dynamics of grey soliton and does not allow a uniform density, which leads to the strong interaction between these hole-like excitations and trap geometry. The Bogoluibov (BG) mode is shown as red dots. In the focusing case (attractive interaction), the energy of these hole-like excitations lies above the single component case, which shows degeneracy (the filled squares) ($\kappa_{ab} = -0.5$). As we decrease the interactions strength ($\kappa_{ab} = -0.2$), the energy gets reduced, but maintains the degeneracy (the filled diamonds). At $\kappa_{ab} = 0$, we retrieve the dispersion behavior of the single component case (the filled upward triangle). As we move towards the defocusing case of a positive sign (repulsive interaction), the inter-species interaction ($\kappa_{ab} = 0.2$) leads to a reduction of energy at each point of high momentum, shown as inverted triangle. If the values of the $\kappa_{ab}$ is increased further, energy becomes negative, that indicates the formation of a bound states, as depicted in the figure by empty circles.

The intriguing aspect of bound state formation between the binary components as manifested in Fig.~\ref{lm-tbec} can be explained in the realm of low energy scattering theory \cite{sakurai}. Here, the effective interaction between the two species, $\kappa_{ab}$ actually relates the transition of the system from weak coupling to the strong coupling regime. In this transition the attractive inter-species interaction $\kappa_{ab}$ points to the weak coupling. The gradual increase in the bare coupling strength results in the effective interaction to be positive, thereby indicating a strong coupling regime which, as in other cases (e.g composite boson formation), favors the bound state formation. This nature is explicated in the figure, as the solitonic energy dips below the zero level. In this context, we would like to point out that such controlled bound state formation may be useful for information storage and retrieval. It is worth noting that grey soliton excitations have non-trivial phase difference; the phase is intensity dependent, whereas in case of bright soliton, it is of kinematic origin. The relative phase can code information, which may be stored in the static configuration, either in the bound state or in the plane static configuration. The phase can be retrieved later by separating them and making them dynamical. Information storage and retrieval in nonlinear media, particularly in active atomic media has already been demonstrated in \cite{gsa}.
%(GSA and TN Dey, they followed a paper of Eberly et al. , two pulse configuration has been used, I think it has been experimentally seen).

\section{Conclusion}
In this article, we study inhomogeneous multi component BEC with special attention to the dispersion mechanism. We present an analytical model to tackle the inhomogeneity and relating them to he homogeneous case. Later, we show that our model is general enough to replicate well known dispersion diagram while we model the interaction in a certain way \cite{liang}. The specific control on the external magnetic field (effectively controlling the interaction) automatically modifies the chirping. Interestingly a generic chirping (with which we started) can enable us to view the hole-like excitation mode experimentally in a small momentum window before getting destroyed. Our study affirms the recent description of the finite lifetime soliton in a trap \cite{gang} in a situation when we employ coherent control on the chirping. The modulated chirping enforces instability for the soliton but it also provides a scope to experimentally observe the Lieb mode in a tiny momentum window. Further, we note that, through our formalism, it is also possible to get back the usual dispersion diagrams akin to the general understanding by means of chirp-engineering. This opens up the possibility for the coherent control and manipulation of these hole-like excitations at low energy, through the temporal modulation of the scattering length, as well as the trap. Additionally, we have shown a wide class of exact solutions for the coupled quasi-one dimensional GP equations in a trap, describing the dynamics of binary condensates. We concentrated on the case where one component possesses a hole-like excitations and  the other being a bright soliton. The formation of bound state in repulsive inter-species interaction is shown and discussed that it is manifested from the low energy scattering theory.

%The nature of this non-uniform BEC profile needs further investigation to see the reliability of local density the

\section*{Acknowledgement}
The Authors acknowledge the fruitful discussions with S. S. Ranjani. PD acknowledges Indian Institute of Science Education and Research, Kolkata, for providing the facilities, where this work has been started. PD also acknowledges the financial support from T\"UB\.{I}TAK-1001, Grant No. 114F170.

%\section*{References}

\end{document}